\documentclass[
amsmath,amssymb,pra,
twocolumn
]{revtex4}
\usepackage{graphicx}% Include figure files
\usepackage{dcolumn}% Align table columns on decimal point
\usepackage{bm}% bold math
\usepackage{hyperref}% add hypertext capabilities
\newcommand{\ket}[1]{\ensuremath{\left| #1 \right>}}
\newcommand{\bra}[1]{\ensuremath{\left< #1 \right|}}
\newcommand{\braket}[2]{\ensuremath{\left< #1 \ \vphantom{#2} \right| 
\left. #2 \vphantom{#1} \right>}}

 \newcommand{\brakety}[2]{\ensuremath{\left( #1 \ \vphantom{#2} \right| 
\left. #2 \vphantom{#1} \right>}}
\newcommand{\kety}[1]{\ensuremath{\left| #1 \right)}}

\newcommand{\sgn}[1]{\ensuremath{\text{sgn}\left( #1 \right) }}
 
\begin{document}
%\colorlet{lightblue}{blue!10}

\preprint{APS/123-QED}

\title{Time Evolution of  Superradiance }% Force line breaks with \\

\author{Colin Rylands} 
\email{rylands@physics.rutgers.edu}
\author{Natan Andrei}
\email{natan@physics.rutgers.edu}
\affiliation{Department of Physics, Rutgers University
Piscataway, New Jersey 08854.
}

\date{\today}% It is always \today, today,

\begin{abstract}
The superradiant behaviour of the Dicke model is examined using the Yudson representation. This is achieved by computing the time evolution of the  mean photon current density and photon number. Extensions of this model including energy splitting and spatial separation are then investigated using this technique.

\end{abstract}

\pacs{Valid PACS appear here}
\maketitle

\section{Introduction}
It is 60 years since Dicke~\cite{Dicke} put forth the notion of cooperative  spontaneous emission. That is, in a system of atoms interacting with a common electromagnetic field each atom cannot be treated in isolation. Rather, they act in a cooperative manner and emit a photon pulse with a higher than expected intensity. This phenomenon is known as superradiance. In particular, for a completely inverted population the transition to the ground state happens on a time scale $\tau\sim 1/M$, $M$ being the number of atoms~\cite{Andreev, Gross}. Dicke's original model considered a  system of two-level atoms coupled to an EM field (treated classically) via electric dipole transition. He showed an enhancement in the probability that a single photon is emitted when the initial state has small $S^z$ quantum number.  Subsequently similar models, in which the system is contained in a cavity and coupled to single mode of a quantum EM field were extensively studied,
\[
H=\omega_ca^{\dag}a+\omega_0S^z+c\left(a^{\dag}S^-+aS^+\right)
\] 
$\omega_c$ and $\omega_0$ being the cavity and transition frequencies respectively. The resonant model $\omega_c=\omega_0$ was solved exactly~\cite{Tavis} and its thermodynamic properties studied. It was found that the single mode and multimode models exhibited a phase transition from normal to superradiance at finite temperature~\cite{Wang,Lieb}. The superradiant phase being characterised by $\left<S^z\right>/M\rightarrow0$, in close analogy with ferromagnetism. Time dependent observables were also calculated~\cite{Scharf} and found  to be oscillatory with the period being sensitive to the number of photons in the initial state. Later, zero temperature phase transitions were also proven when the system is either collectively driven or damped.~\cite{ZeroTpt1,ZeroTpt2}

More recently there has been renewed interest in this field. Models coupling photons to two-level atoms or, in more modern parlance, fluxoniums, are of great interest due to the applications in quantum information theory and quantum computing. Experimentally, great advancements have seen the observation of phase transitions for the first time and cooperative effects in cold atomic gases (see~\cite{garraway} and references therein) as well as the successful construction of a superradiant laser~\cite{SLaser}.
\begin{figure}
\centering
\includegraphics[trim = 20mm 0mm 0mm 0mm,width=0.5\textwidth]{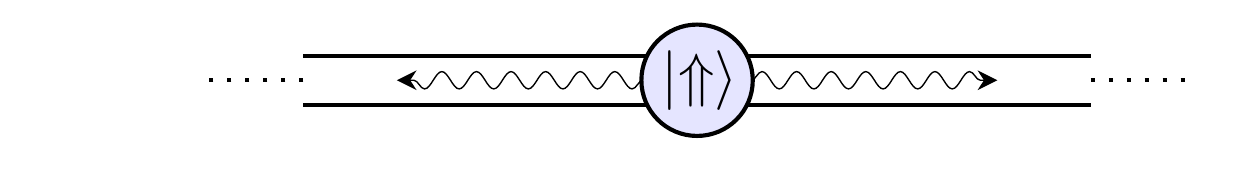}

\caption{A schematic for the system~\eqref{HNC} under consideration. An ensemble of two-level atoms sits at the centre of a 1D  waveguide. The transition from the completely excited to ground state is acompanied by a superradiant photon pulse.} \label{schematic}
\end{figure}
In this article observables of an effective 1-dimensional model, \eqref{HNC}, of superradiance are calculated. This model differs from those described above in that it  includes the full spectrum of the EM field and is  situated in a 1D waveguide of infinite size rather than a cavity. Accordingly no phase transitions of the type already expanded upon are expected. Through the exact time evolution of these observables, using an initial state of excited atoms, superradiance is explicitly shown (there is no oscillatory behaviour for the reasons just stated). It is found that exciting $N$ out a system of $M$ two-level atoms and preparing them in a state characterised by  spin $s\leq\frac{M}{2}$, the superradiance is given by
\begin{eqnarray}\nonumber
\left<\hat{N}_{\text{p}}\right>_t&=&N(1-e^{-cN(1+2s-N)t})\\\nonumber
\partial_t\left<\hat{N}_{\text{p}}\right>_t&=&cN^2(1+2s-N)e^{-cN(1+2s-N)t} 
\end{eqnarray}
$\hat{N}_{\text{p}}$ being the photon number operator. In particular note that when the system initially has a complete population inversion, $s=\frac{M}{2}$, $N=M$ the transition rate is $\propto M^2$, this being a hallmark of the cooperative nature of superradiance.

This question is then investigated in two other variations. The first allows for varying transition frequencies of the atoms, known as inhomogeneous broadening, while the second includes spatial separation of the atoms with or without broadening.
\section{ An Effective 1D Hamiltonian}
Here an effective 1D field theory of superradiance based on the Dicke model is considered. It is a non-chiral version of that obtained by Yudson and Rupasov~\cite{Yudson1, Rupasov} via reduction from the full 3D model of superradiance. The system described is that of left and right moving photons (with the speed of light set to unity) interacting with a collection of $M$ two level atoms located at the origin. The Hamiltonian is 
\begin{align}\label{HNC}\nonumber
H_{\text{nc}}=\int\left(ib^{\dag}_L(x)\partial_xb_L(x)-ib^{\dag}_R(x)\partial_xb_R(x)\right)\\
-\sqrt{c/2}\left( S^+(b_L(0)+b_R(0))+S^-(b^{\dag}_L(0) +b^{\dag}_R(0)\right)
\end{align}
where $b_L^{\dag}(x)$, $b_R^{\dag}(x)$ create left and right moving photons, respectively, at the point $x$, $c$ is the positive coupling constant and $S^+=\sum_{i=1}^Ms^+_i$ the raising operator for the atomic system with $s_i^+$ being the raising operators of individual 2-level atoms.  As all atoms are located at the same point, $x=0$,  one may unfold  the Hamiltonian  in the standard manner for single impurity models. Using the transformation to a Weyl basis
\begin{align}
b(x)=\frac{1}{\sqrt{2}}(b(x)_R+b_L(-x)) \\
b_0(x)=\frac{1}{\sqrt{2}}(b(x)_R-b_L(-x))
\end{align}
one obtains
\begin{eqnarray}\nonumber
H_{nc}&=& -i\int dx\,  b^{\dag}(x)\partial_{x}b(x)-\sqrt{c}\left( S^+b(0)+S^-b^{\dag}(0) \right)\\
&-&i\int dx\,  b_0^{\dag}(x)\partial_{x}b_0(x) 
\end{eqnarray}
where the field  $b^{\dag}_0(x)$ is a decoupled  photon field,  while the field $b^{\dag}(x)$ creates a chiral photon at $x$ which  interacts with the system of two-level atoms. 
Actually it is a combination of photonic  modes  propagating symmetrically left and right  with respect to the impurity.  In the following, initial states used to compute observables will be purely atomic and so the non-interacting photons will not contribute. The superradiance is therefore  described by the field $b(x)$ and is  governed by,
\begin{equation}\label{H}
H= -i\int dx\,  b^{\dag}(x)\partial_{x}b(x)-\sqrt{c}\left( S^+b(0)+S^-b^{\dag}(0) \right).
\end{equation}
This Hamiltonian  commutes with the excitation number, 
\begin{equation} \label{N}
\hat{N}=\int dx \,\rho (x) +\left(S^z+M/2\right)\,,\,\,\,\rho (x)=b^{\dag}(x)b(x)
\end{equation}
being equal to the number of photons and excited atoms and $S^2$ the total spin.

The Hamiltonian, \eqref{H},  is  integrable and its  highest weight  maximal spin  eigenstates in the subspace $N$  were given by Rupasov and Yudson.  The completion of diagonalisation to include lower spin states is due to A. Culver \cite{Culver}. The highest weight and maximal spin  eigenstates in the subspace $N$ are  given by,
\begin{eqnarray}\nonumber
\ket{\vec{\lambda}}&=& \frac{1}{\left(2 \pi\right)^{\frac{N}{2}}N!^{\frac{1}{2}}}\int d^Nx \prod_{i<j}\left(1-\frac{2 i c \theta(x_i-x_j)}{\lambda_i-\lambda_j+i c}\right) \\\label{L}
&&\times\prod_{j=1}^Ne^{i\lambda_j x_j}f(\lambda_j,x_j)r^{\dag}(\lambda_j,x_j)\ket{0}.
\end{eqnarray}
Here
\begin{equation}
r^{\dag}(\lambda_j,x_j)=b^{\dag}(x_j)-\frac{\sqrt{c}}{\lambda_j}S^+
\end{equation}
which ensures \eqref{L} are also eigenstates of \eqref{N} and $S^2$, while
\begin{align}\label{f}
f(\lambda_j,x_j)=\frac{\lambda_j-icM/2\,\text{sgn}(x_j)}{\lambda_j+icM/2}
\end{align}
describes the phase shift acquired by the photon as it crosses the the M-atom system, and
$1-\frac{2 i c \theta(x_i-x_j)}{\lambda_i-\lambda_j+i c}$ captures the photon-photon interactions induced by the impurity.
The set of rapidities  $\vec{\lambda}=(\lambda_1,\dots\lambda_N)$  parametrises the solution. The ground state $\ket{0}$ contains only unexcited atoms, $S^2\ket{0}=\frac{M}{2}\left(\frac{M}{2}+1\right)\ket{0}$, $S^z\ket{0}=-\frac{M}{2}\ket{0}$. Note \eqref{f} has a pole at $-iM c/2$ signalling a bound state of a photon and unexcited atom. The corresponding eigenvalues are 
\begin{eqnarray}
H\ket{\vec{\lambda}}&=&\sum_{i=1}^{N}\lambda_i\ket{\vec{\lambda}}\,,\,\,\hat{N}\ket{\vec{\lambda}}=N\ket{\vec{\lambda}},\\
S^2\ket{\vec{\lambda}}&=&\frac{M}{2}\left(\frac{M}{2}+1\right)\ket{\vec{\lambda}}.
\end{eqnarray}
The states $\ket{\vec{\lambda}}$ form a complete set on the $M+1$ dimensional subspace spanned by $(S^+)^n\ket{0}\,,\,n=0,1,\dots,M$, the completely symmetric part of the atomic Hilbert space. Using these solutions, eigenstates with lower $S^2$ eigenvalues are obtained via the following construction:
Denote by $\mathcal{O}_{i_1,\dots, i_{\alpha}}$ the operator which acts on the vacuum to create a Bethe state of the form \eqref{L} but only for atoms $i_1\dots i_{\alpha}$. This is then an eigenstate of $H$ for $M=\alpha$ with  atoms labelled  $i_1\dots i_{\alpha}$. It follows that
\begin{align}
\mathcal{O}_{3,\dots ,M}\frac{\left(s_1^+-s_2^+\right)}{\sqrt{2}}\ket{0}
\end{align}
will be an eigenstate of \eqref{H}, \eqref{N}  and $S^2$ with the eigenvalue of the later being $\frac{M}{2}\left(\frac{M}{2}-1\right)$. The subspace with this total spin is degenerate and these other states are obtained by taking suitable linear combinations of $\mathcal{O}_{i_i,\dots ,i_{M-2}}\left(s_{i_{M-1}}^+-s_{i_{M}}^+\right)\ket{0}$. Lower multiplets are obtained by repeating this procedure. The singlet state does not interact with the photons and is already an eigenstate of \eqref{H}.

We now turn to the time evolution  of superradiance, when the system is quenched from an initial state of of excited $N, N\le M$ atoms.  An arbitrary initial state can be represented by utilising the  contour integral method developed by Yudson ~\cite{Yudson1}~\cite{Andrei}. This replaces the usual formula 
\begin{align}
\ket{\Psi_0}=\sum_{\vec{\lambda}}\ket{\vec{\lambda}}\braket{\vec{\lambda}}{\Psi_0}
\end{align}
with
\begin{align}
\ket{\Psi_0}=\int_{\Gamma}d\vec{\lambda}\ket{\vec{\lambda}}\brakety{\vec{\lambda}}{\Psi_0}
\end{align}
where the integration is carried out over contours in the complex $\lambda$ plane which depend upon the initial state as well as the S-matrices present in the Bethe state and
\begin{multline}
\kety{\vec{\lambda}}=\frac{N!}{2\pi^{N/2}}\int d^Nx \\
 \theta(x_{1}\!\geq\!\dots\!\geq\!x_N\!)\prod_{i=1}^Ne^{i\lambda_i x_i}f(\lambda_i,x_i)r^{\dag}(\lambda_i,x_i)\ket{0}.
\end{multline}
The superradiance of \eqref{H} is examined by computing the time evolved photon current density, $j(z)= v \rho(z)$ (with $v=1$, in present units) and the photon number, $\hat{N}_{\text{p}}$,
\begin{eqnarray} \label{j}
\left< j(z)\right>_t=\left<\rho(z)\right>_t&=&\bra{\Psi_0}e^{-i H t}\rho (z)e^{i H t}\ket{\Psi_0}\\
\label{NP}
\left< \hat{N}_{\text{p}}\right>_t&=&\int dz \left<\rho(z)\right>_t
\end{eqnarray}
where $\ket{\Psi_0}$ is an arbitrary atomic state. Presently, only states in the symmetric atomic Hilbert space are considered,   $\ket{\Psi_0}=\left( \frac{(M-N)!}{M!N!} \right)^{1/2}\!(S^+)^N \!\ket{0}$. The first equality in \eqref{j} is due to the choice of units and the chirality of \eqref{H}. The relation of \eqref{j},\eqref{NP} to the corresponding quantities for the non-chiral system are
\begin{eqnarray}
\left< j_R(z)\right>_t&=&\frac{1}{2}\left< j(z)\right>_t|_{c\rightarrow c/2},\\
\left< j_L(z)\right>_t&=&-\frac{1}{2}\left< j(-z)\right>_t|_{c\rightarrow c/2},\\
\left< \hat{N}_{\text{R\,p}}\right>_t=\left< \hat{N}_{\text{L\,p}}\right>_t&=&\frac{1}{2}\left< \hat{N}_{\text{p}}\right>_t|_{c\rightarrow c/2}.
\end{eqnarray}
Using the Yudson representation, the chosen initial state, expanded in terms of Bethe states is 
\begin{align}
\ket{\Psi_0}=\left[\frac{M!}{(M-N)!N!}\right]^{1/2}\int_{\Gamma}\frac{d^N\lambda}{2\pi^{N/2}}\prod_{i=1}^N\frac{-\sqrt{c}}{\lambda_i-i c M/2}\ket{\vec{\lambda}}.
\end{align}
The contours required are given by $\Gamma=\gamma_1\otimes\dots\otimes\gamma_N$,
\begin{align}
 \text{Im} \gamma_{i+1}-\text{Im}\gamma_i>c \\
c M/2> \text{Im}\gamma_i>-c M/2
\end{align}

\begin{figure}
  \centering
    \includegraphics[width=0.45\textwidth]{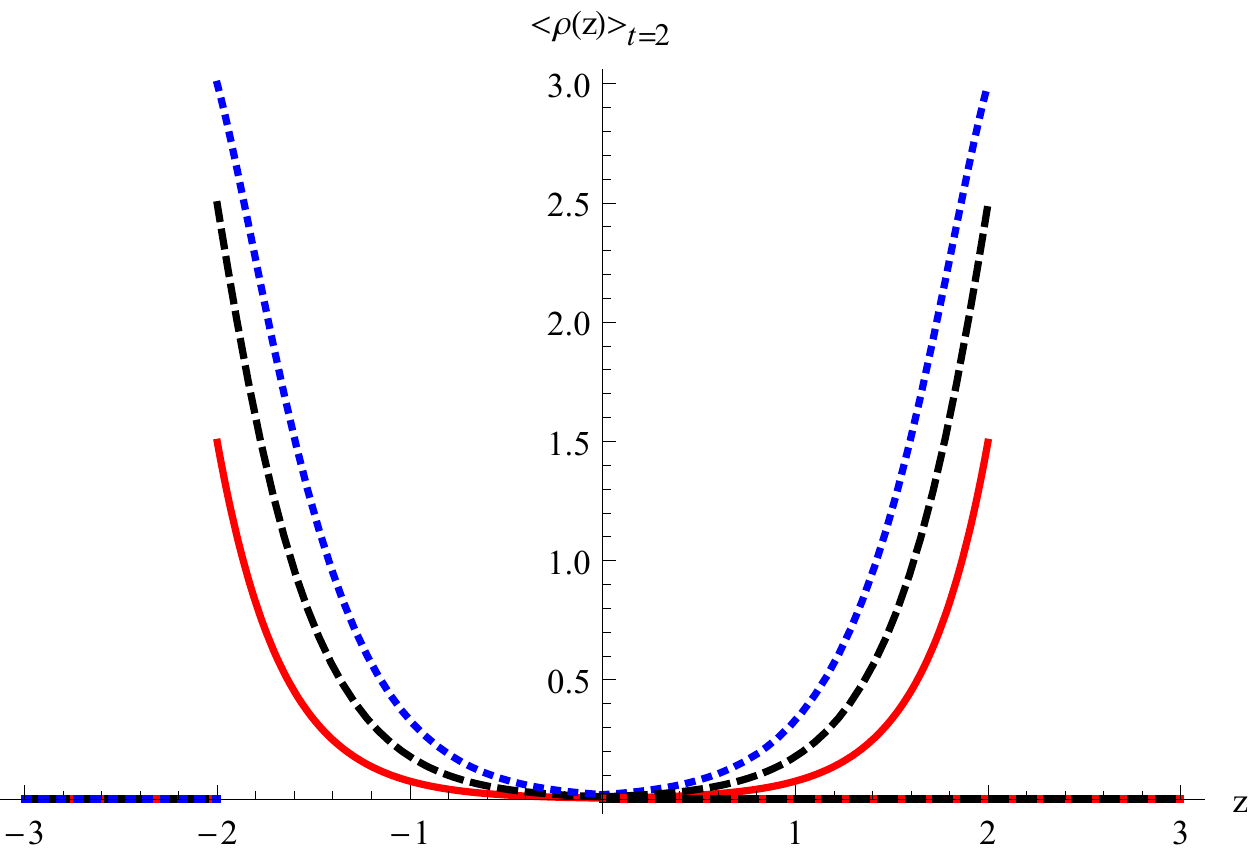}
 \includegraphics[width=0.45\textwidth]{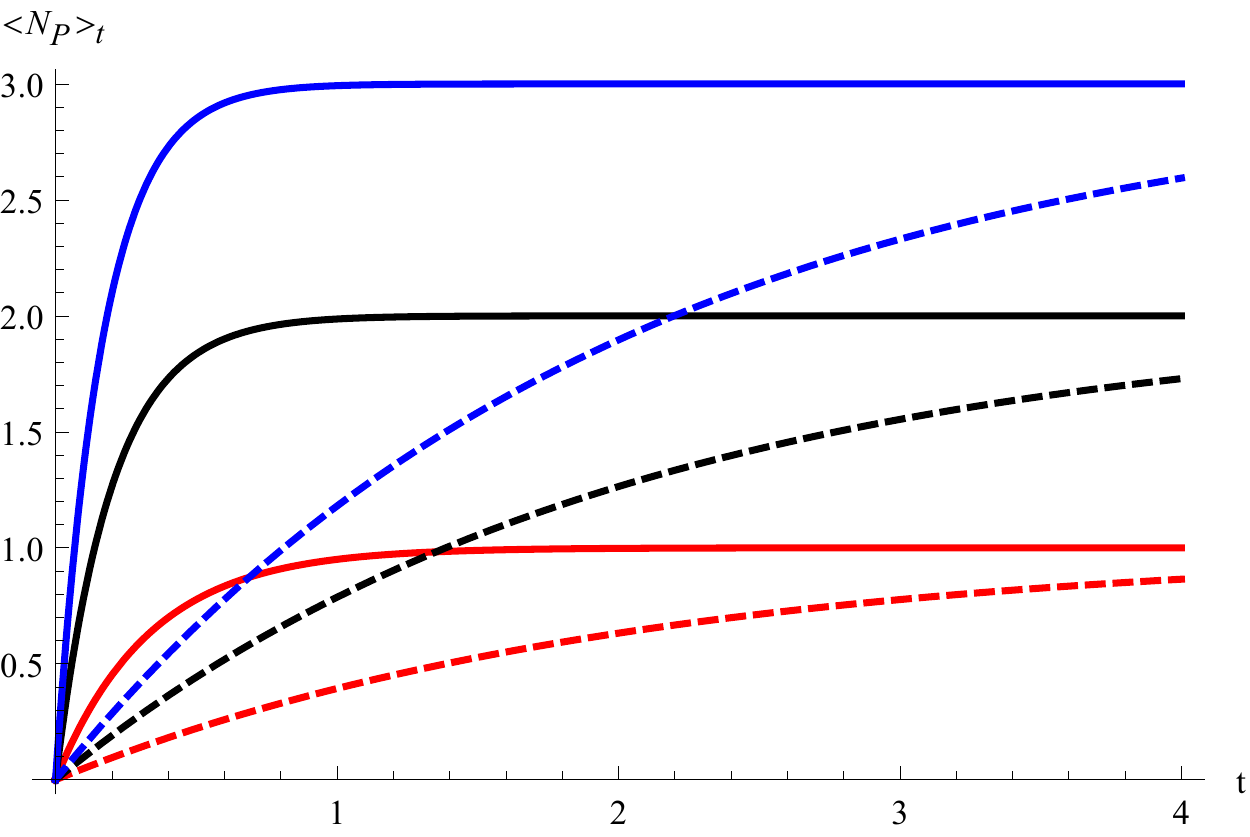}
  \caption{(colour online)(a) $\left<j_{L,R}(z)\right>_t$ at $t=2$  for $c=1$, $M=6$, $N=3$ (blue, dotted), $N=2$ (dashed) and $N=1$ (red) for the non-chiral model.  There is no delay time evident in  $\left<j_{L,R}(z)\right>_t$, the peak of the emitted pulse lies on the light cone, $t=\pm z$. This is attributed to the atoms being all located at the origin. (b) The total photon number $\left< \hat{N}_{\text{R\,p}}\right>_t+\left< \hat{N}_{\text{L\,p}}\right>_t$ for $c=1$, $M=6$, $N=3$ (blue), $N=2$ (black) and $N=1$ (bed) as well the corresponding expressions ignoring cooperative effects, \eqref{NN}, (dashed) as a function of time  (These always lie below the values for \eqref{SR}). One sees that when cooperative effects are taken into account the number operator approaches its long time value much faster.} \label{1}
\end{figure}

The evolution of this state is computed \cite{Yudson2} by including a factor of $e^{-i \lambda t}$ and closing the contours in the lower half plane. Starting with the the $\lambda_1$ integration only distinct poles are picked up. The result is
\begin{align}\label{psit}
\begin{split}
\ket{\Psi_0(t)}=\left( \frac{N!}{(M-N)!}\right)^{1/2}e^{\frac{-c N(1+M-N)t}{2}}\sum_{m=0}^{N}(i\sqrt{c})^{N-m} \\
\times \!\frac{(M-n)!}{m!}\int dx_{m+1}\dots dx_N \theta(0\!<\!x_{m+1}\!<\!\dots\!<\!x_N\!<\!t) \\
e^{\frac{c}{2}\sum_{j=m+1}^N(M+2-2j)x_j} (S^+)^m\!\!\! \prod_{i=m+1}^N b^{\dag}(x_i)\ket{0}.
\end{split}
\end{align}
The desired quantities can now be evaluated.  
\begin{multline}\label{rho}
\left<\rho(z)\right>_t=\!\frac{N!}{(M-N)!}e^{-c N(1+M-N)t}\sum_{m=0}^{N-1}c^{N-m}\frac{(M-m)!}{m!} \\
\times \int dx_{m+1}\dots dx_N \theta(0\!<\!x_{m+1}\!<\!\dots\!<\!x_N\!<\!t) \\
e^{c\sum_{j=m+1}^N(M+2-2j)x_j}\left(\sum_{j=m+1}^N\delta(z-x_j)\right)
\end{multline}
Here, only the trivial permutation in the sum $\sum_{\mathcal{P}}\prod_i \delta \left(x_i-x'_{\mathcal{P}_i}\right) $ has  contributed due to the Heaviside function. The integrals in \eqref{rho} can be explicitly evaluated by separating out the Heaviside function  \'{a} la
\begin{multline}
 \theta(0\!<\!x_1\!<\!\dots\!<\!x_2\!<\!t)=\\
\theta(x_1)\theta(x_2)\theta(t)\theta(x_2-x_1)\theta(t-x_1)\theta(t-x_2)
\end{multline}
The expression obtained for $\left< \rho(z)\right>_t$ vanishes outside of the casual region $z<t$ due to the
\[\sum_{j=m+1}^N\delta(z-x_j) \theta(0\!<\!x_{m+1}\!<\!\dots\!<\!x_N\!<\!t)\]
 which will produce an overall $\theta(z<t)$. In addition, it vanishes for $z\leq 0$ as should be the case for system comprising of just right moving photons. The exponential decay of \eqref{rho} depends on both $N$ and $M$ and as will be explained shortly, this a signifier of superradiance. Lastly, note that each term in the sum involves $N-m-1$ integrals resulting in a factor of $c^{-N+m+1}$ which leaves $\left<\rho(z)\right>_t\propto c$. Figure~\ref{1} shows  $\left< \rho(z)\right>_t$ for different values of the parameters. It is evident that the peak intensity is located at $t=z$, i.e. there is no delay time. The absence of a delay time can be attributed to the atoms all being at the origin.

The effect of cooperation amongst atoms is more transparent in the quantity  \eqref{NP}. To this end, first note that as $\ket{\Psi_0}$ is normalised $1=\braket{\Psi_0\!}{\!\Psi_{0}}=\braket{\Psi_0(t)\!\!}{\!\Psi_{0}(t)}$, thus
\begin{multline}\label{one}
1\!=\!\frac{N!}{(M-N)!}e^{-c N(1+M-N)t}\sum_{m=0}^{N}c^{N-m}\frac{(M-n)!}{m!} \\
\times \int dx_{m+1}\dots dx_N \theta(0\!<\!x_{m+1}\!<\!\dots\!<\!x_N\!<\!t) \\
e^{c\sum_{j=m+1}^N(M+2-2j)x_j}.
\end{multline}
To obtain $\left< \hat{N}_{\text{p}}\right>_t$ one integrates \eqref{rho} over $z$. This effectively makes the $\sum_{j=m+1}^N\delta(z-x_j)$ term redundant and one almost reproduces $N$ copies of the identity \eqref{one} apart from the $m=N$ term which is purely atomic, therefore
\begin{eqnarray}\label{SR}
\left<\hat{N}_{\text{p}}\right>_t&=&N(1-e^{-cN(1+M-N)t}) \\ \label{SRT}
\partial_t\left<\hat{N}_{\text{p}}\right>_t&=&cN^2(1+M-N)e^{-cN(1+M-N)t} .
\end{eqnarray}
Comparing this to the naive result ignoring the cooperative nature of the emitters 
\begin{eqnarray}\label{NN}
\left< \hat{N}'_{\text{p}}\right>_t&=&N(1-e^{-ct}) \\
\partial_t\left< \hat{N}'_{\text{p}}\right>_t&=&cNe^{-ct}
\end{eqnarray}
one sees that the superradiant system enjoys an enhancement of its coupling constant to $c_{SR}=cN(1+M-N)$, $cM\leq c_{SR}\leq cM(M+1)/2$. It is at a maximum when half of the atoms are initially excited, this is in agreement with the result of Dicke~\cite{Dicke}. In particular when the initial state is completely excited the transition rate to the atomic ground state is $cM^2$. Alternatively one can see the time scale, $\tau$, for transition to the atomic ground state is significantly decreased from $1/c$ to $1/c_{SR}$ and  for the completely excited initial state $\tau = 1/Mc$. The model \eqref{H} includes the full photon spectrum, however the dispersion of emitted photons approaches a Lorentzian sharply peaked about the transition frequency of the atoms~\cite{Markov} at long times\footnote{While the Hamiltonian,~\eqref{H}, is dependent on the transition frequency, $\omega_0$, only through the coupling constant $c$, this being essentially the single atom transition amplitude, it should be understood that all frequencies (wavevectors) be measured from $\omega_0$ ($k_0$) (See~\cite{Yudson1}). For instance, the distribution for an emitted photon from the first excited atomic state at long time is $\frac{\frac{c_{SR}}{2}}{\pi(k^2+(\frac{c_{SR}}{2})^2)}$. This should be read as $\frac{\frac{c_{SR}}{2}}{\pi((k-k_0)^2+(\frac{c_{SR}}{2})^2)}$. }. Hence in terms of the intensity $I \sim M \omega_0/\tau=cM^2\omega_0$. Thus the two most common descriptors of superradiance are shown. Emergence of the coupling enhancement can be traced back to the pole structure of the photon-impurity S-matrix. The collection of atoms was treated as a single impurity with pole strength $-icM/2$. Superradiant  behaviour is evident even when $N=1$,
\begin{eqnarray}\nonumber
\left<\rho(z)\right>_t&=&cM\theta(z\!<t)e^{-cM(t-z)}\\\label{rho1}
\left<\hat{N}_{\text{p}}\right>_t&=&1-e^{-cMt}
\end{eqnarray}
providing a simple test of whether a system is superradiant or not.

The above calculations proceed in a similar fashion if the an initial state with $s<M/2$ were chosen. The resulting expressions also involve an enhancement of the coupling constant to $c_{SR}=cN(1+2s-N)$. Therefore, using a completely general atomic initial state specified by $s\leq\frac{M}{2}$, $N\leq 2s$
\begin{align}
S^2\ket{\Psi_0}=s\left(s+1\right)\ket{\Psi_0},\,\,\,\hat{N}\ket{\Psi_0}=N\ket{\Psi_0}
\end{align}
the relevant expectation values are given by
\begin{eqnarray}\label{Ns}\nonumber
\left<\hat{N}_{\text{p}}\right>_t&=&N(1-e^{-cN(1+2s-N)t})\\
\partial_t\left<\hat{N}_{\text{p}}\right>_t&=&cN^2(1+2s-N)e^{-cN(1+2s-N)t} 
\end{eqnarray}
and
\begin{multline}\label{rhos}
\left<\rho(z)\right>_t=\!\frac{N!}{(2s-N)!}e^{-c N(1+2s-N)t}\sum_{m=0}^{N-1}c^{N-m}\frac{(2s-m)!}{m!} \\
\times \int dx_{m+1}\dots dx_N \theta(0\!<\!x_{m+1}\!<\!\dots\!<\!x_N\!<\!t) \\
e^{c\sum_{j=m+1}^N(2s+2-2j)x_j}\left(\sum_{j=m+1}^N\delta(z-x_j)\right).
\end{multline}
Again, the strongest effect of cooperation is felt when $N=s$ for $M$ even or $N=s+1/2$ for $M$ odd. Furthermore, the coupling enhancement persists for $N=1$ provided $s>1$.

\section{Generalisations}
\begin{figure}
  \centering
    \includegraphics[width=0.45\textwidth]{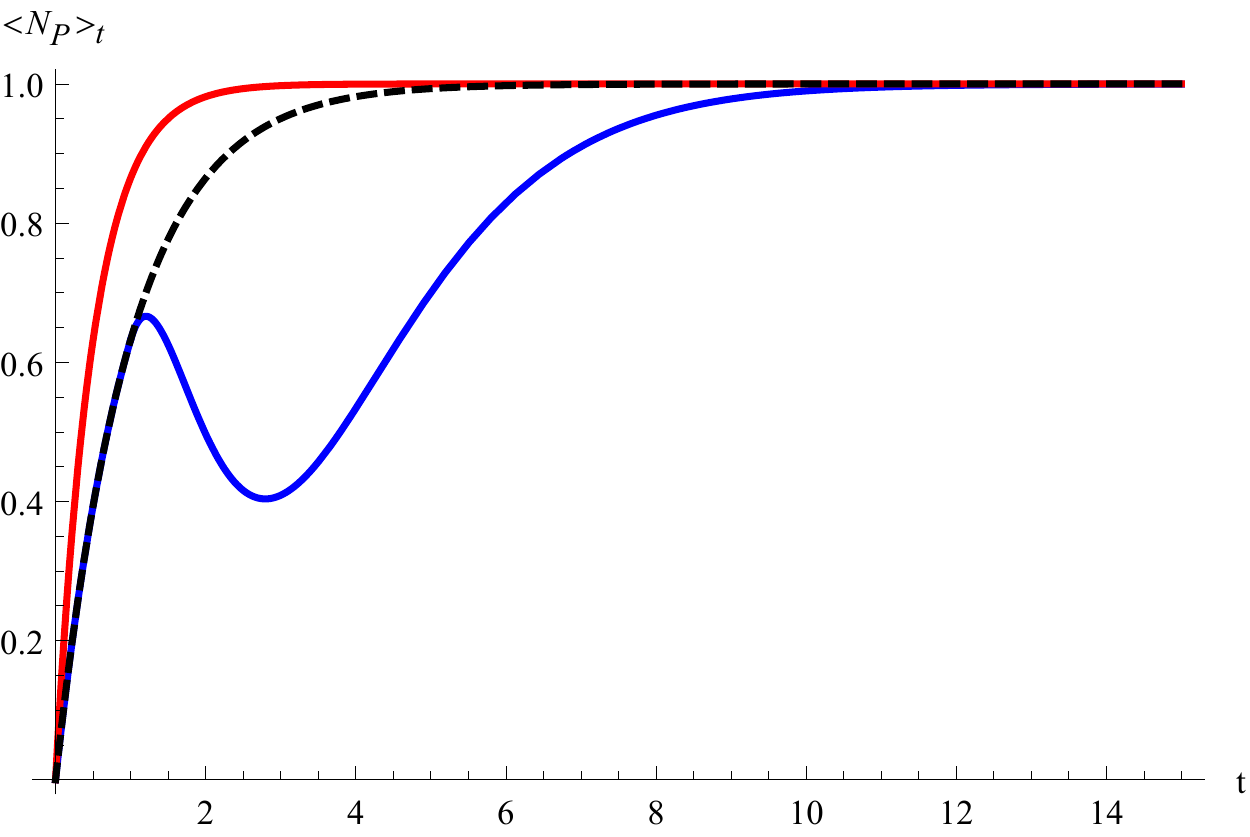}
\includegraphics[width=0.45\textwidth]{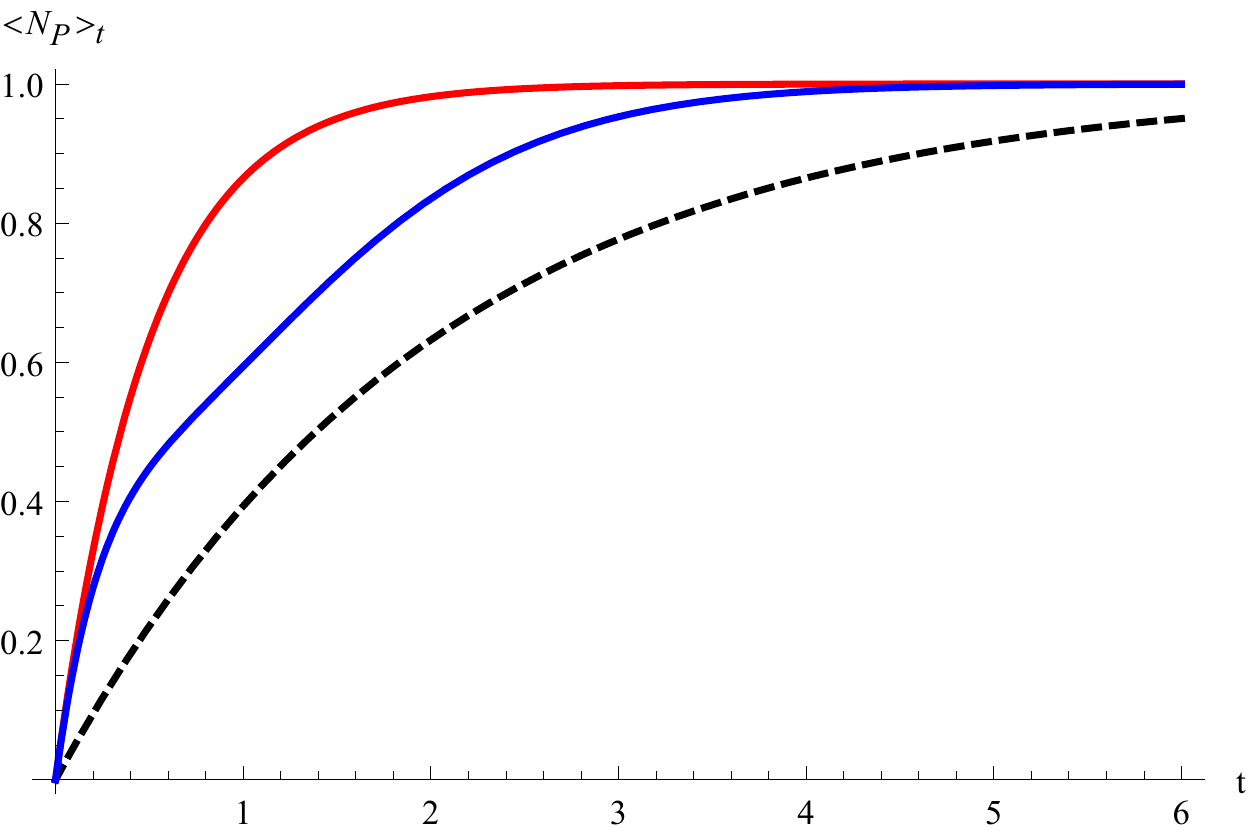}
  \caption{(colour online) (a) $\left<\hat{N}_{\text{p}}\right>_t$  for two spatially separated atoms the first of which is excited (blue), two Dicke type atoms (red) and two non cooperative atoms (dashed), $c=1,\,r=1$. The asymptote is reached slower when there is spatial separation than for cooperation or non-cooperation although it agrees with the later when $t<r$.  (b) The case of two enegry splittings, $M=8,\,m=4,\,N=1,\Delta=1$ (blue)  again the expression for cooperative and non-cooperative atoms are included also. The curve with splitting lies between both, indicating the prescence of some cooperation when there is energy splitting.}\label{2Impurities}
\end{figure}

Possible extensions of~\eqref{H} which are also integrable include allowing atoms to have varying transition frequencies~\cite{Yudson1} known as inhomogeneous broadening, to be spatially separated~\cite{Yudson2} or both~\cite{Yudson3}. To incorporate inhomogeneous broadening into the model an energy splitting term, $\sum_{k=1}^M\Delta_k(s^z_k+1/2)$, is added to \eqref{H}. The eigenstates of this model can also be constructed via the Bethe ansatz and are given by \eqref{L} with
\begin{eqnarray}\label{fin}
f(\lambda_j,x_j)&=&\frac{1-i c\, \text{sgn}(x_j) /2 \left(\sum_{k=1}^M(\lambda_j-\Delta_k)^{-1}  \right)}{1+i c/2 \left(\sum_{k=1}^M(\lambda_j-\Delta_k)^{-1}  \right)} \\\nonumber
r^{\dag}(\lambda_j,x_j)&=&b^{\dag}(x_j)-\sqrt{c} \delta(x_j)\left(\sum_{k=1}^M(\lambda_j-\Delta_k)^{-1}s^+_k  \right).
\end{eqnarray}
The energy splittings should be regarded as variations from $\omega_0$, thus the restriction $\sum_{k=1}^M\Delta_k=0$ is imposed. As should be expected \eqref{fin} has as many poles as different values of $\Delta_k$. The Bethe states can now discern between the different types of atom and when all are different they form a complete set on the full atomic Hilbert space. To make the problem more tractable one can specialise to the case of only two types of atom. Denoting the two splittings $\Delta$ and $\Delta'$ and allowing for $m$ primed atoms, one has $\Delta'=-\frac{M-m}{m}\Delta$. Accordingly, $f(\lambda_j,x_j)$ is given by
\begin{align}
\frac{(\lambda_j-\Delta)(\lambda_j+\frac{M-m}{m}\Delta)-i c M \text{sgn}(x_j)(\lambda_j+(\frac{M}{m}-2)\Delta)}{ (\lambda_j-\Delta)(\lambda_j+\frac{M-m}{m}\Delta)+i c M (\lambda_j+(\frac{M}{m}-2)\Delta)}
\end{align}
with poles at 
\begin{multline}\label{Lpm}
\lambda^{\pm}=-\frac{icM}{4}\mp\frac{iMc}{4\sqrt{2}}\times\\
\sqrt{\sqrt{\left(1-\frac{4\Delta^2}{m^2c^2}\right)^2+\left(\frac{4\Delta}{Mc}\left(\frac{M}{m}-2\right)\right)^2}+\left(1-\frac{4\Delta^2}{m^2c^2}\right)} \\
\pm\frac{Mc}{4\sqrt{2}}\times\\
\sqrt{\sqrt{\left(1-\frac{4\Delta^2}{m^2c^2}\right)^2+\left(\frac{4\Delta}{Mc}\left(\frac{M}{m}-2\right)\right)^2}-\left(1-\frac{4\Delta^2}{m^2c^2}\right)}\\
-\frac{\Delta}{2}\left(\frac{M}{m}-2\right).
\end{multline}
Initial states comprising an excited atomic system can be represented using the Yudson representation;
\begin{eqnarray}\nonumber
(S^+)^N\ket{0}=(-1)^N \left(\frac{c^{N/2}(M-m)!}{(M-N-m)!}\right) \\
\times \int_{\Gamma}\frac{d^N\lambda}{2\pi^{N/2}}\prod_{i=1}^N\frac{(\lambda_i+\frac{M-m}{m}\Delta)}{(\lambda_i-\lambda^{*+})(\lambda_i-\lambda^{*-})}\ket{\vec{\lambda}},\\\nonumber
(S'^+)^N\ket{0}=(-1)^N\left(\frac{c^{N/2}m!}{(m-N)!}\right)\quad\quad\,\,\\
\times\int_{\Gamma}\frac{d^N\lambda}{2\pi^{N/2}}\prod_{i=1}^N\frac{(\lambda_i-\Delta)}{(\lambda_i-\lambda^{*+})(\lambda_i-\lambda^{*-})}\ket{\vec{\lambda}}.
\end{eqnarray}
Here the contours must be taken so as to lie between the poles closest to the real axis, $\text{Im}\lambda^{*-}> \text{Im}\gamma_i> \text{Im}\lambda^-$ and $\text{Im} \gamma_{i+1}-\text{Im}\gamma_i>c$. This poses a problem for arbitrary $N$, one may not be able to fit the required number of contours between these points. A best case is when there are equal numbers of primed and unprimed atoms and $1\leq \frac{4\Delta}{Mc}$ whence the $M/2$-fold excited state can be represented. The first excited state can, however,  always be represented, with the contour simply running along the real axis. In fact, this is all that is required for the present purposes. If there are equal numbers of primed and unprimed atoms there is considerable simplification to 
\begin{align}
\lambda^{\pm}=-\frac{icM}{4}\left(1\pm\sqrt{1-\frac{16\Delta^2}{M^2c^2}}\right)
\end{align}
with $N=1,\,m=M/2$ the time dependent photon part is given by,
\begin{multline}
\frac{e^{-iHt}S^+\ket{0}}{\sqrt{M/2}}=-\left( \frac{M}{2ca^2}\right)^{\frac{1}{2}}\int dx \theta(x\!<t)e^{-\frac{Mc}{4}(t-x)}\\
\times\left((\lambda^+\!\pm\Delta)e^{-\frac{aMc}{4}(t-x)}-(\lambda^-\!\pm\Delta)e^{\frac{aMc}{4}(t-x)}\right) b^{\dag}(x)\ket{0}
\end{multline}
where $a=\sqrt{1-\left(\frac{4\Delta}{Mc}\right)^2}\neq0$ and the upper sign corresponding to unprimed atoms the lower to primed ones. If $a=0$ i.e. $Mc=4\Delta$ the poles coincide and the result is 
\begin{multline}
\frac{e^{-iHt}S^+\ket{0}}{\sqrt{M/2}}=i\sqrt{\frac{Mc}{2}}\int dx  \theta(x\!<t)e^{-\frac{Mc}{4}(t-x)} \\
\times\left(\left(-\frac{Mc}{4}\pm\Delta\right)\left(-i(t-x)+1\right)\right) b^{\dag}(x)\ket{0}.
\end{multline}
The current density is given by,
\begin{widetext}
\begin{multline}
 \left<\rho(z)\right>_t= \begin{cases}\frac{1}{16a^2} e^{-\frac{Mc}{2}(t-z)} \theta(z\!<t)
\times \left(\left(\left(\frac{cM}{4}\right)^2(1+a^2)+\Delta^2\right)\cosh{\left(cMa(t-z)/2\right)}
+\left(\frac{cM}{4}\right)^2(1+a^2)\right.\\
-\Delta^2+\left.\frac{M^2c^2a}{8}\sinh{\left(cMa(t-z)/2\right)}\right) &
\text{if}\,Mc\neq4\Delta\\
\frac{Mc}{2} e^{-\frac{Mc}{2}(t-z)} \theta(z\!<t)\left(1-\frac{cM}{2}(t-x) 
+\left(\left(\frac{cM}{4}\right)^2+\Delta^2\right)(t-x)^2\right) &
\text{if}\,Mc=4\Delta.
\end{cases}
\end{multline}
\end{widetext}
Comparing this with \eqref{rho1}, some superradiant behaviour is evident but it is lessened by the energy splitting giving $c_{SR}=cM/2$ (See Fig.~\ref{2Impurities}). This occurs due to the dependence of \eqref{Lpm} on $M$. For the system with $M$ different splittings $f(\lambda_j,x_j)$ is
\begin{align}
\frac{\prod_{k=1}^{M}(\lambda_j-\Delta_k)-i\frac{c}{2}\sgn{x_j}\sum_{l=1}^M\prod_{k\neq l}^M(\lambda_j-\Delta_k)}{\prod_{k=1}^{M}(\lambda_j-\Delta_k)+i\frac{c}{2}\sum_{l=1}^M\prod_{k\neq l}^M(\lambda_j-\Delta_k)}.
\end{align}
The single pole in \eqref{f} has split into $M$ different poles each of which is independent of the total number of atoms. Therefore no superradiance is expected and is recovered in part only when multiple poles coincide.
When considering atoms with spatial separation (known as the MacGillivray Feld model~\cite{MF}) and inhomogeneous broadening the interaction term in \eqref{H} is replaced by 
\begin{align}
-\sqrt{c}\sum_{i=1}^M \left( s_i^+b(r_i)+s_i^-b^{\dag}(r_i) \right)+\sum_{k=1}^M\Delta_k(s^z_k+1/2)
\end{align}
$r_i$ being the position of the $i^{th}$ atom The eigenstates are given by \eqref{L} ~\cite{Yudson3} and 
\begin{eqnarray}\label{fmf}
f(\lambda_j,x_j)&=&\prod_{k=1}^M\frac{\lambda_j-\Delta_k-i\frac{c}{2}\sgn{x_j-r_k}}{\lambda_j-\Delta_k+i\frac{c}{2}}\\
r^{\dag}(\lambda_j,x_j)&=&b^{\dag}(x_j)-\sqrt{c}\left(\sum_{k=1}^M\frac{\delta (x_j-r_k)s_k^+}{\lambda_j-\Delta_k}\right).
\end{eqnarray}
One should now immediately recognise that there is no superradiance  in this model for any value of splittings. The individual treatment of the atoms present in \eqref{fmf} means that the atoms are not initially correlated and due to the chirality of the photons no correlations can be produced via photon exchange. Thus no cooperative emission can occur. Indeed one can proceed as before, obtaining for the $n^{\text{th}}$ atom initially excited and no broadening
\begin{multline}
\left<\rho(z)\right>_t=ce^{-c(t+r_n-z)}\left(\sum_{k=0}^{M-n}\sum_{m=0}^k\theta(r_{n+k}\!<\!x\!<r_{k+n+1})\right.\\
\left.\theta (x\!<\!t+r_n)\frac{\left(c(x-r_n-t)\right)^m}{m!} {k \choose m}\right)^2.
\end{multline}
As anticipated the exponential decay is independent of the number of atoms.
\section{Conclusion}
In conclusion, the observables $\left<j(z)\right>_t$ and  $\left< \hat{N}_{\text{p}}\right>_t $ have been calculated from an arbitrary initial atomic state. The results show explicitly  the cooperative emission produced in the Dicke model. It exhibits the characteristic $\propto M^2$ transition rate of superradiance but lacks the time delay. Two extensions of the Dicke model were then examined. The first, allowing for inhomogeneous broadening, was found to exhibit cooperative emission whose effect was reduced compared to the case of no splitting. The second model incorporated spatial separation and broadening  and it was found, by considering the pole structure of the photon impurity S-matrix, that no superradiance is evident. 

\acknowledgements{The authors would like to thank Adrian Culver for many helpful and constructive discussions.  We are  grateful to  M. Imada, H. Tureci and V.I Yudson  for useful comments. This research was supported by NSF grant
DMR 1410583.
}

\bibliography{mybib}

\begin{thebibliography}{10}

\bibitem{Dicke}
R.~H. Dicke,
\newblock Phys. Rev. {\bf 93}, 99 (1954).

\bibitem{Andreev}
A.~V. Andreev, V.~I. Emel’yanov, and Y.~A. Il’inskii,
\newblock Physics-Uspekhi {\bf 23}, 493 (1980).

\bibitem{Gross}
M.~Gross and S.~Haroche,
\newblock Physics Reports {\bf 93}, 301  (1982).

\bibitem{Tavis}
M.~Tavis and F.~W. Cummings,
\newblock Phys. Rev. {\bf 170}, 379 (1968).

\bibitem{Wang}
Y.~K. Wang and F.~T. Hioe,
\newblock Phys. Rev. A {\bf 7}, 831 (1973).

\bibitem{Lieb}
K.~{Hepp} and E.~H. {Lieb},
\newblock Annals of Physics {\bf 76}, 360 (1973).

\bibitem{Scharf}
G.~Scharf,
\newblock Annals of Physics {\bf 83}, 71  (1974).

\bibitem{ZeroTpt1}
R.~Puri, S.~Lawande, and S.~Hassan,
\newblock Optics Communications {\bf 35}, 179  (1980).

\bibitem{ZeroTpt2}
S.~V. Lawande, R.~R. Puri, and S.~S. Hassan,
\newblock Journal of Physics B: Atomic and Molecular Physics {\bf 14}, 4171
  (1981).

\bibitem{garraway}
B.~M. Garraway,
\newblock Philosophical Transactions of the Royal Society A: Mathematical,
  Physical and Engineering Sciences {\bf 369}, 1137 (2011).

\bibitem{SLaser}
J.~G. Bohnet {\em et~al.},
\newblock Nature , 78–81 (2012).

\bibitem{Yudson1}
V.~Rupasov and V.~Yudson,
\newblock Zh {\`E}ksp Teor Fiz {\bf 87}, 1617 (1984).

\bibitem{Rupasov}
V.~Rupasov and V.~Yudson,
\newblock Zh. Eksp. Teor. Fiz {\bf 86}, 825 (1984).

\bibitem{Culver}
A.~Culver,
\newblock Private communication.

\bibitem{Andrei}
D.~Iyer and N.~Andrei,
\newblock Phys. Rev. Lett. {\bf 109}, 115304 (2012).

\bibitem{Yudson2}
V.~Yudson,
\newblock Zh. Eksp. Teor. Fiz {\bf 88}, 1757 (1985).

\bibitem{Markov}
A.~A. Makarov and V.~I. Yudson,
\newblock Phys. Rev. A {\bf 89}, 053806 (2014).

\bibitem{Yudson3}
V.~I. {Yudson},
\newblock Physics Letters A {\bf 129}, 17 (1988).

\bibitem{MF}
J.~C. MacGillivray and M.~S. Feld,
\newblock Phys. Rev. A {\bf 14}, 1169 (1976).

\end{thebibliography}

\bibliographystyle{h-physrev}

\end{document}